\documentclass[11pt]{article}

\usepackage[a4paper,margin=1in]{geometry}
\usepackage{amsmath,amssymb,bm}
\usepackage{mathtools}
\usepackage{graphicx}
\usepackage{booktabs}
\usepackage{placeins}
\usepackage{hyperref}
\usepackage{cite}
\hypersetup{colorlinks=true,linkcolor=blue,citecolor=blue,urlcolor=blue}

\newcommand{\dd}{\mathrm{d}}
\newcommand{\ReLU}{\mathrm{ReLU}}
\newcommand{\SiLU}{\mathrm{SiLU}}
\newcommand{\elu}{\mathrm{ELU}}

\title{Reconstructing Meson Parton Distributions from Mellin Moments\\
with Artificial Neural Networks}
\author{%
\begin{tabular}[t]{p{0.92\textwidth}}
\centering
Pietro Dall'Olio$^{1}$\thanks{\texttt{pietro.dallolio@lyon.edu}}\,,
Cristian Del Gobbo$^{1,2}$\thanks{\texttt{delgobbo@uark.edu}}\,,
Kh\'epani Raya$^{3,4}$\thanks{\texttt{khepani.raya@dci.uhu.es}}\,,
Jos\'e Rodr\'iguez-Quintero$^{3}$\thanks{\texttt{jose.rodriguez@dfaie.uhu.es}}
\\[0.6em]
{\small $^{1}$Division of Mathematics and Science, Lyon College, Batesville, AR 72501, USA}\\
{\small $^{2}$Department of Mathematical Sciences, University of Arkansas, Fayetteville, AR 72701, USA}\\
{\small $^{3}$Dpto. Ciencias Integradas, Centro de Estudios Avanzados en Fis., Mat. y Comp.,
Fac. Ciencias Experimentales, Universidad de Huelva, E-21071 Huelva, Spain}\\
{\small $^{4}$Departamento de Actuar\'ia, F\'isica y Matem\'aticas, Universidad de las Am\'ericas Puebla,
San Andr\'es Cholula, 72820 Puebla, M\'exico}
\end{tabular}%
}
\date{\today}

\begin{document}
\maketitle

\begin{abstract}
We address the inverse problem of reconstructing a meson parton distribution
function (PDF) or parton distribution amplitude (PDA) at a fixed scale from a
finite set of Mellin moments. At finite moment order this map is ill posed in
the Hadamard sense: the moments do not determine a unique, stable pointwise
$x$-dependence, and small data perturbations can produce large changes in the
reconstruction. We regularize the problem by restricting the admissible
function class through constraints justified by first principles---support on
$x\in[0,1]$, endpoint vanishing, unit normalization, and, for the pion, the
exact $x\leftrightarrow 1-x$ symmetry---and by complementing these hard
constraints with differentiable penalties that suppress the spurious multiple
maxima and central concavity changes typical of ill-posed inversions. We
parameterize the unknown function with two interchangeable neural cores, a
multilayer perceptron (MLP) and a Kolmogorov--Arnold network (KAN), and find
that the KAN reconstructs the target distributions more accurately, most
markedly in the asymmetric (kaon) case. We attribute this advantage to the
locality of the B-spline basis underlying the KAN, which localizes corrections
during moment matching and damps the global oscillations to which the MLP is
prone. Benchmarks against synthetic targets, using both ten and five input
moments, support these conclusions.
\end{abstract}

\section{Introduction}
Parton distribution functions (PDFs) and parton distribution amplitudes (PDAs)
encode, respectively, the longitudinal momentum structure of hadrons in
inclusive processes and the leading-twist light-cone valence structure relevant
to exclusive ones~\cite{Ellis:1996mzs,LepageBrodsky,
EfremovRadyushkin}. In several nonperturbative frameworks---most notably lattice
QCD and Dyson--Schwinger/Bethe--Salpeter approaches (DSE/BSE), see, e.g.,~\cite{Raya:2024ejx,Chang:2013pq,Xu:2023bwv,Xu:2026rtu,Cui:2021mom,Cui:2022bxn,Yin:2023dbw,Lu:2023yna,Xu:2024nzp,Ding:2019qlr,Cui:2020tdf,Xu:2026sjy,RQCD:2019osh,Joo:2019bzr,Alexandrou:2021mmi,Gao:2022iex}---one does not access the
full $x$-dependence directly but rather a finite set of Mellin moments at a
scale $\mu$,
\begin{equation}
  M_n(\mu)\equiv \int_0^1 x^{n}\, q(x,\mu)\,\dd x,\qquad n=0,1,\dots,N,
  \label{eq:mellin}
\end{equation}
where $q$ denotes generically a valence PDF or a PDA. The goal of this work is
to reconstruct $q(x,\mu)$ on $x\in[0,1]$ from $\{M_n(\mu)\}_{n=0}^N$ with as
little functional bias as possible.

Reconstructing $q$ from finitely many moments is an inverse problem of
intrinsically ill-defined character. Hadamard well-posedness requires
existence, uniqueness, and stability~\cite{Hadamard,TikhonovArsenin,EHN};
finite-$N$ moment inversion generically violates at least two of the three: infinitely many
functions share the same first $N$ moments, and small moment perturbations may
induce large variations in the reconstructed $x$-profile~\cite{ShohatTamarkin}.
In practice, flexible reconstructions tend to develop unphysical oscillations,
extra bumps (multiple maxima), or curvature changes near the peak that are not
supported by meson-distribution phenomenology, precisely because nothing in
the finite moment data itself rules them out.

A natural response is to postulate a rigid parametric form---a Beta-like ansatz
with a handful of shape parameters---and fit only those parameters to the
moments. This controls the pathology at the cost of introducing a different,
harder-to-quantify bias: the reconstruction can only ever look like a member of
the assumed family, however inappropriate that family turns out to be for the
true $q(x)$. Entropy-based approaches offer an alternative means of regularizing the reconstruction, see, e.g.,~\cite{Chen:2024dhz}. However, as with a prescribed functional ansatz, the resulting distribution is not determined by the moments alone, but also reflects the additional criterion used to select among the admissible solutions. We instead let a flexible neural network stand in for $q(x)$ and
ask how little, rather than how much, needs to be assumed. This is where the
choice of network matters: an ordinary multilayer perceptron (MLP), with
enough width or depth, can approximate any continuous function on $[0,1]$
arbitrarily well, a guarantee formalized by the universal approximation
theorem~\cite{Hornik,Cybenko}. The Kolmogorov--Arnold network (KAN), the
second architecture used in this work, rests on an analogous but distinct
guarantee, the Kolmogorov--Arnold representation theorem~\cite{Kolmogorov},
which shows that every continuous multivariate function decomposes exactly
into a finite sum of compositions of continuous univariate functions. In
neither case is the theorem itself the algorithm we use---practical networks
of finite size only approximate the guarantee---but both theorems play the
same conceptual role here: they certify that, before any constraint is
imposed, the network is in principle capable of representing the true $q(x)$,
whatever its shape. The bias in our reconstruction is therefore not the
implicit bias of an insufficiently expressive functional form, but exactly the
explicit, physically motivated bias we choose to impose and nothing more. This
is the precise sense in which we call the fit unbiased.

That bias is imposed through a short list of constraints, detailed in Sec.~2,
which encode what is actually known about $q(x)$ independently of the moment
data: the kinematic support and normalization of the distribution, its
behavior at $x=0,1$, the exchange symmetry of the pion, and---on somewhat
different, statistical grounds---the expectation that $q(x)$ has a single
maximum and no reversal of curvature near it. We enforce a constraint by
hard-wiring it into the network architecture whenever an exact closed-form
realization is available, and fall back to a soft, differentiable penalty
added to the loss only when it is not. This mirrors a broader distinction in
the physics-informed neural network literature~\cite{Raissi}, where physical
laws are typically imposed as soft penalties on the loss (e.g.\ residuals of a
governing PDE); we prefer hard constraints wherever possible, since a
property enforced by construction cannot be traded off against the
moment-matching term during training and holds exactly for every value of
$\theta$. The same combination---a flexible neural core dressed with a
physically motivated, hard-wired prefactor---was pioneered in high-energy
phenomenology by the NNPDF collaboration's neural-network parametrization of
parton distributions, in which a preprocessing factor $x^{\alpha}(1-x)^{\beta}$
multiplying the network output controls the Regge and quark-counting endpoint
behavior while the network itself remains free to fit everything
else~\cite{NNPDF2002,NNPDF30}; the endpoint prefactors of Sec.~3.3 follow the
same logic. Neural-network parametrizations of this kind have since been
extended beyond collinear PDFs to generalized parton distributions (GPDs),
where the same nonparametric, bias-minimizing philosophy has been used to
model the additional skewness and momentum-transfer dependence while
respecting polynomiality and positivity~\cite{Dutrieux2022,DallOlio2024}, and,
most recently, to learn parton distributions directly from lattice-QCD matrix
elements within a single neural-network framework spanning collinear PDFs and
zero-skewness GPDs~\cite{Chu2026}. Section 2 specifies and justifies each
constraint; Sec.~3
introduces the two network cores used to realize them, together with the
elements of spline theory needed to follow the KAN construction (collected in
Appendix~\ref{app:bsplines}).
We find that the KAN systematically
improves the reconstructions, especially for asymmetric distributions such as
the kaon's, and we argue in Sec.~3 that the locality of its B-spline
activations is the likely reason.

\section{Physical constraints}
Define the forward operator mapping a function to its first $N$ moments,
\begin{equation}
  \mathcal{A}: q\mapsto \bm{M}\in\mathbb{R}^N,\qquad
  (\mathcal{A}q)_n=\int_0^1 x^{n}q(x)\,\dd x ,\quad n=1,\dots,N.
  \label{eq:forward}
\end{equation}
At finite $N$ the operator $\mathcal{A}$ has an infinite-dimensional nullspace:
there exist infinitely many $\delta q\neq 0$ with $\mathcal{A}\,\delta q=\bm 0$,
so the reconstruction is non-unique, and it is unstable even after
non-uniqueness is reduced. Restricting the admissible function class through
properties that hold independently of the moment data is the standard remedy,
and it is most useful when each restriction can be tied as directly as
possible to the physics of the quantity being reconstructed rather than to
convenience.

The variable $x$ is a light-cone momentum fraction and is kinematically
confined to $[0,1]$. The unit-area condition
\begin{equation}
  \int_0^1 q(x)\,\dd x = 1
  \label{eq:norm}
\end{equation}
has a direct meaning in both settings we consider: a leading-twist PDA is
normalized to the meson decay constant, i.e.\ to unity once the constant is
factored out~\cite{LepageBrodsky,EfremovRadyushkin}, while a single-flavor
valence PDF integrates to the valence-quark number, again unity for a meson.

A second constraint concerns the behavior at the two endpoints of this same
interval. For a leading-twist PDA the vanishing $q(0)=q(1)=0$ is rigorous: the
asymptotic distribution is $\varphi_{\mathrm{as}}(x)=6\,x(1-x)$, and
corrections retain the linear endpoint suppression~\cite{LepageBrodsky,
EfremovRadyushkin}. For a valence-quark PDF, the endpoint condition $q(1)=0$ is likewise required~\cite{BrodskyFarrar,FarrarJackson,Berger:1979du}: $x=1$ corresponds to the struck quark carrying the entire meson momentum, leaving the remaining partons with zero momentum; consequently, the corresponding endpoint configuration has vanishing probability. Consistent with this behavior, at resolution scales comparable to or below the proton mass, quark-counting rules predict the large-$x$ behavior
$q(x)\sim(1-x)^{\beta}$, with $\beta=2$ in the partonic limit of QCD~\cite{BrodskyFarrar,FarrarJackson,Berger:1979du}. As the resolution scale increases, QCD evolution leads to a larger value of $\beta$~\cite{Dokshitzer:1977sg,Gribov:1972ri,Lipatov:1974qm,Altarelli:1977zs}. This pattern is supported by lattice QCD, data-driven methods, continuum approaches, and symmetry-driven studies (see, e.g., Refs.~\cite{Sufian:2020vzb,Xu:2023bwv,Lu:2023yna,Xu:2026rtu,Xu:2024nzp,Chen:2024dhz,Cui:2021mom,Cui:2022bxn,Ding:2019qlr,Cui:2020tdf,Yin:2023dbw,Chang:2020kjj,Hecht:2000xa,Holt:2010vj,Wang:2025usl}). The behavior at $x\to 0$ is more delicate: Regge phenomenology
gives $q(x)\sim x^{-\alpha}$ for valence distributions~\cite{BallNoceraRojo},
which need not vanish---it can rise while remaining integrable ($\alpha<1$)---and the
constraint fails outright for sea and gluon distributions, which we do not
consider here. However, we choose to work at the so-called hadron scale, at which the fully dressed valence quarks constitute the sole degrees of freedom, encoding all properties of the hadron, and hence share the entire hadron momentum; see, e.g., Refs.~\cite{Xu:2026rtu,Xu:2023bwv,Xu:2024nzp,Lu:2023yna,Cui:2021mom,Cui:2022bxn,Ding:2019qlr,Cui:2020tdf,Yin:2023dbw}. Then, at this scale, the two-body valence structure of the meson relates the quark $q(x)$ and antiquark $\bar{h}(x)$ distributions as:
\begin{equation}
\bar{h}(x)=q(1-x)\,.
\end{equation}
Therefore, the large-$x$ behavior of the quark distribution determines the small-$x$ behavior of the antiquark distribution, and vice versa. The endpoint condition $q(1)=0$ thus implies $\bar{h}(0)=0$, while $\bar{h}(1)=0$ sets $q(0)=0$. In other words, at the hadron scale $q(0)=q(1)=0$ is also a rigorous constraint for a valence PDF, as $x=0$ also corresponds to a configuration of vanishing probability.

Furthermore, for those mesonic systems made of a valence quark and antiquark of the same flavor ($h=q$), such as {\it quarkonia}, or specifically for the pion in the light-flavor sector and isospin-symmetric limit, one is left with a valence PDF at the hadron scale which obeys
\begin{equation}
  q(x)=q(1-x),
  \label{eq:symmetry}
\end{equation}
implying as a third constraint the exchange symmetry under $x \leftrightarrow 1-x$, which reflects the equal sharing of momentum between the valence quark and antiquark of equal mass. For the PDA the same follows from charge
conjugation and G-parity, which force all odd Gegenbauer (equivalently,
odd-about-$x=\tfrac12$ Mellin) moments to vanish~\cite{LepageBrodsky,RQCD:2019osh}. Notably, the above symmetry prescription, together with the approximately monopole behavior of the pion electromagnetic form factor, is sufficient to generate a smooth endpoint behavior for the pion valence PDF~\cite{Wang:2025usl}. The symmetry is broken
for the kaon by SU(3) flavor breaking: the heavier strange quark carries more
momentum on average, producing a skewed distribution with a nonzero first
Gegenbauer moment $a_1^{K}\neq 0$~\cite{ChernyakZhitnitsky,RQCD:2019osh}. Taken
together, these first three constraints are generically justified for PDAs (the
exchange symmetry, of course, only for the pion and quarkonia), whereas for PDFs
some of them---particularly the vanishing at $x\to0$ and the exchange
symmetry---should be regarded as a consequence of evaluating the distribution at
the hadron scale. In the pion case, an exact isospin-symmetric limit is also
required.

A further constraint, not fixed by symmetry or endpoint behavior but no less
physically motivated for mesonic ground states, concerns the overall shape of $q(x)$ in between: we
expect it to rise monotonically from the origin to a single maximum and fall
monotonically thereafter, without additional bumps, and without a reversal of
curvature (an inflection point) in the region around that maximum; see, e.g.,
Refs.~\cite{Raya,Raya:2024ejx,Xu:2026rtu}. Unlike the
constraints above, this one is not a matter of exact kinematics or an exact
symmetry, and it is worth spelling out on what grounds it can be expected. For
the leading-twist PDA the argument is close to a theorem. The scale dependence
of a PDA is governed by the ERBL evolution equation~\cite{Lepage:1979zb,Efremov:1979qk,Lepage:1980fj}, whose eigenfunctions are
the Gegenbauer polynomials $C_n^{3/2}(1-2x)$ built on the same $x(1-x)$ weight
as the asymptotic form $\varphi_{\mathrm{as}}(x)$ above, with eigenvalues
(one-loop anomalous dimensions) that grow monotonically with
$n$~\cite{LepageBrodsky,EfremovRadyushkin,ChernyakZhitnitsky}. Every Gegenbauer moment beyond the
lowest is therefore driven toward zero under evolution, increasingly so for
higher $n$, so that the PDA is pulled toward the single-humped asymptotic
shape and any admixture capable of generating a secondary maximum or a central
inflection is perturbatively suppressed---consistent with the small,
rapidly decreasing Gegenbauer coefficients found in modern lattice and
continuum extractions~\cite{RQCD:2019osh,Chang:2013pq}. Early QCD sum-rule
analyses pointed in the opposite direction: Chernyak and Zhitnitsky proposed
a double-humped pion PDA, $\varphi_{\mathrm{CZ}}(x)=30\,x(1-x)(1-2x)^2$,
corresponding to a large second Gegenbauer coefficient
$a_2=2/3$~\cite{ChernyakZhitnitsky}; this form is disfavored by the later
lattice and continuum results just cited. A second, more general argument
covers the valence-PDF interpretation as well, for which no analogous
evolution equation singles out a preferred basis: a smooth, single-humped
shape is the generic, least-structured solution to a moment-constrained
reconstruction problem, which is the content of the maximum-entropy
principle~\cite{Jaynes}. Under that principle, the function compatible with a
finite set of moments that adds no structure beyond what the data require is
the maximum-entropy density, which for a low-order moment constraint is itself
unimodal---matching only a mean and a variance, for instance, already selects
a Gaussian-like single-humped profile. The same logic underlies the
maximum-entropy method used to reconstruct hadronic spectral functions from a
finite, noisy set of Euclidean-correlator moments in lattice QCD~\cite{AHN},
where spurious multi-peak structures are likewise recognized as reconstruction
artifacts rather than genuine features. We stress that, away from the PDA
case, this is a physically motivated modeling choice rather than a derived
result, consistent with---but not proved by---the phenomenology of
single-flavor valence distributions; Sec.~4 turns this expectation into two
explicit, differentiable penalties.

\section{Network architectures}
We construct $q(x)$ from a univariate network core $f(x;\theta)$, evaluated
pointwise on $x\in[0,1]$, and then wrap it with the algebraic operations of
Sec.~3.3 that enforce the hard constraints of Sec.~2 exactly. Two functionally
interchangeable cores are used and compared: a multilayer perceptron (MLP) and
a Kolmogorov--Arnold network (KAN). As discussed in Sec.~1, both are, in the
appropriate limit, universal representations of a one-dimensional continuous
function; they differ in \emph{where} their learnable flexibility lives, on
the nodes for the MLP and on the edges for the KAN, and this difference is
what Sec.~3.2 argues is responsible for the KAN's better behavior under moment
matching.

\subsection{Multilayer perceptron}
A feedforward MLP builds $f(x;\theta)$ by alternating affine maps
$W(z)=Az+b$ with a fixed, elementwise nonlinearity $\sigma$: the input is
combined linearly into a layer of units, each unit is passed through $\sigma$,
and the resulting vector becomes the input to the next layer. With $L$ layers,
\begin{equation}
  f_{\mathrm{MLP}}(x;\theta)=\big(W^{(L)}\!\circ \sigma\circ W^{(L-1)}\circ
  \cdots\circ\sigma\circ W^{(1)}\big)(x),
  \label{eq:mlp}
\end{equation}
which for a single hidden layer of width $H$ reduces to the familiar
one-layer form $f=b^{(2)}+\sum_{j=1}^{H}w^{(2)}_j\,
\sigma_{\elu}(w^{(1)}_j x+b^{(1)}_j)$. We use the ELU activation,
\begin{equation}
  \sigma_{\elu}(z)=
  \begin{cases}
    z, & z\ge 0,\\
    \lambda\,(e^{z}-1), & z<0,
  \end{cases}
  \label{eq:elu}
\end{equation}
with fixed $\lambda>0$, and do not impose positivity on the output. The
parameters $\theta$ are the collection of all weights $A^{(\ell)}$ and biases
$b^{(\ell)}$; the nonlinearity $\sigma$ itself is never trained, so all of the
network's flexibility resides in how these linear weights combine and
recombine a fixed nonlinear building block.

\subsection{Kolmogorov--Arnold network}
Since their introduction, KANs have found a range of applications in
physics~\cite{Liu:2025kanscience}: in high-energy physics, to event
classification at the LHC~\cite{Erdmann:2024unt,Abasov:2024hyq}, to the
analysis of lattice-QCD data on the quark--antiquark flux tube~\cite{Kou:2024hzd},
and to the symbolic rediscovery of the Gell-Mann--Okubo mass
relations~\cite{He:2026tev}; in nuclear physics, to binding-energy
prediction~\cite{Liu:2024dnj} and as variational many-body wave
functions~\cite{Bedaque:2025cas}; to holographic inverse
problems~\cite{Dai:2025dir,Jeong:2025omu,Kim:2026mll}; and, further afield, in
cosmology and gravitational-wave astronomy~\cite{Cerardi:2025ato,Liu:2025snl}.
To our knowledge, they have not yet been applied to the reconstruction of
parton distributions.

The KAN~\cite{KAN} departs from this picture by placing a trainable
nonlinearity on every edge of the network, in the form of a spline, and
summing only at the nodes. We use here only the property of splines that
matters for the construction and for the argument below---that a spline can be
written as a linear combination of \emph{compactly supported} basis functions,
so that a single coefficient only ever affects the curve locally---and refer
to Appendix~\ref{app:bsplines} for the explicit definitions.

Formally, a KAN layer with $n_{\mathrm{in}}$ inputs and $n_{\mathrm{out}}$
outputs is a matrix of learnable univariate functions
$\Phi_\ell=\{\phi_{\ell,j,i}\}$, one per edge, and each node value is simply
the sum of its already-transformed incoming signals,
\begin{equation}
  x_{\ell+1,j}=\sum_{i=1}^{n_\ell}\phi_{\ell,j,i}(x_{\ell,i}),
  \qquad
  f_{\mathrm{KAN}}(x;\theta)=\big(\Phi_{L-1}\circ\Phi_{L-2}\circ\cdots\circ\Phi_0\big)(x).
  \label{eq:kan}
\end{equation}
This construction is motivated by the Kolmogorov--Arnold representation
theorem~\cite{Kolmogorov}, which guarantees that any continuous
$f:[0,1]^{n}\to\mathbb{R}$ can be written through univariate functions and
addition alone,
\begin{equation}
  f(x_1,\dots,x_n)=\sum_{q=1}^{2n+1}\Phi_q\!\Big(\sum_{p=1}^{n}\phi_{q,p}(x_p)\Big),
  \label{eq:ka-theorem}
\end{equation}
in the same way that the universal approximation theorem motivates
Eq.~\eqref{eq:mlp}: a KAN of depth two and the prescribed width could in
principle realize Eq.~\eqref{eq:ka-theorem} exactly, and deeper, wider KANs
are used in practice for greater flexibility. Each edge function $\phi$
combines a fixed residual activation with a trainable spline built from the
B-spline basis $\{B_i(z)\}$ of Appendix~\ref{app:bsplines},
\begin{equation}
  \phi(z)=w_b\,\SiLU(z)+w_s\sum_{i}c_i\,B_i(z),
  \qquad \SiLU(z)=\frac{z}{1+e^{-z}},
  \label{eq:kan-edge}
\end{equation}
where $\SiLU$ is a single fixed function, shared by every edge and nonzero
across the whole domain, while each $B_i$ is one element of a fixed \emph{local}
basis, nonzero only over a small window of $z$ (Appendix~\ref{app:bsplines});
the two terms of Eq.~\eqref{eq:kan-edge} therefore play different roles even
though both come with a trainable weight. The spline order and knot grid
underlying $\{B_i\}$ are fixed, and the trainable parameters $\theta$ on every
edge are the coefficients $c_i$ together with the two scalars $w_b,w_s$. The
architectural contrast with the MLP is now precise: the MLP fixes a
nonlinearity on its nodes and learns only linear weights on its edges, while
the KAN learns the nonlinearity itself, edge by edge, and only sums on its
nodes; for the present one-dimensional problem the network reduces, layer by
layer, to a composition of learnable spline curves.

The compact support of the B-spline basis means that fitting a KAN to the
moments adjusts $f(x;\theta)$ \emph{locally} in $x$: a gradient step that
improves the fit in one region of the domain leaves the spline unchanged in
regions far from it. The MLP has no analogous structure---every weight enters
activations with global support in $x$---so a gradient step aimed at
correcting the fit in one region can perturb it everywhere, which is precisely
the mechanism that seeds the domain-spanning spurious oscillations
characteristic of ill-posed moment inversion. This picture is consistent with
our empirical finding that the KAN's advantage over the MLP is most visible for
the asymmetric kaon distribution, whose peak is displaced from $x=\tfrac12$ and
whose reconstruction therefore benefits most from being able to sculpt one side
of the domain without disturbing the other.

\subsection{Hard-constraint wrappers}
The constraints of Sec.~2 are imposed on the network output, independently of
the core. For the symmetric case (in the following, we consider $q_\pi$ for a pion PDF as a paradigmatic symmetric case, although the same applies generically to any quarkonium distribution), we symmetrize and subtract the endpoint
value,
\begin{equation}
  f_{\mathrm{sym}}(x)=\tfrac12\big(f(x)+f(1-x)\big),\qquad
  \tilde q_\pi(x)=f_{\mathrm{sym}}(x)-f_{\mathrm{sym}}(0),
  \label{eq:pion-wrap}
\end{equation}
which guarantees $\tilde q_\pi(0)=\tilde q_\pi(1)=0$ and exact
$\tilde q_\pi(x)=\tilde q_\pi(1-x)$ \emph{without} an explicit prefactor. For the
asymmetric case (here specialized to the kaon), we multiply by a fixed endpoint prefactor,
\begin{equation}
  \tilde q_K(x)=x(1-x)\,f(x),
  \label{eq:kaon-wrap}
\end{equation}
matching the linear PDA endpoint suppression; the prefactor can be generalized
to a (learnable) $x^{\alpha}(1-x)^{\beta}$ to accommodate the
Brodsky--Farrar exponent $\beta=2$ of valence PDFs, in the same spirit as the
NNPDF preprocessing factor mentioned in Sec.~1. In both cases the final
distribution is normalized by construction,
\begin{equation}
  q(x)=\frac{\tilde q(x)}{\int_0^1\tilde q(t)\,\dd t}.
  \label{eq:hardnorm}
\end{equation}

\section{Loss function and shape regularization}
Neither core, as defined above, is ever asked to match the training data
directly. The data are the moments $\{M_n^{\mathrm{tar}}\}$---integrals of
$x^{n}q(x)$ over $[0,1]$---while the network only ever produces a pointwise
value $f(x;\theta)$, and hence $q(x;\theta)$ after the wrapping of Sec.~3.3, at
whatever $x$ it is evaluated. Turning these pointwise values into moments
therefore requires a numerical integration step, and this step has to happen
\emph{before} every gradient-descent update, not just once at the end of
training: at each iteration we evaluate $q(x_i;\theta)$ on a fixed mesh
$\{x_i\}_{i=0}^{M}$ covering $[0,1]$ and combine these samples with fixed
quadrature weights into estimates $M_n^{\mathrm{pred}}(\theta)$ of the moments
(and, by the same construction, of the normalization integral of
Eq.~\eqref{eq:hardnorm}); only these estimates, not the pointwise values
themselves, enter the loss below\footnote{As an unlimited number of predicted moments can be numerically estimated from a distribution evaluated on the fixed mesh, the full sequence $\{M_m^{\mathrm{pred}}(\theta)\}_{m=0}^{\infty}$ also provides a formal, unique representation of the network's output, $q(x;\theta)$. Thus, formally, the network can also be seen to deliver, at each iteration, the moments $\{M_m^{\mathrm{pred}}(\theta)\}_{m=0}^{N'}$ from the training data $\{M_n^{\mathrm{tar}}\}_{n=0}^{N}$ for any $N'>N$; and the loss minimization should be understood as bringing the $N'-N$ remaining moments closer to the unknown values for $q(x)$, to the same extent that this distribution is reproduced by $q(x;\theta)$ over its full support.}. Because the mesh and quadrature weights are
fixed, this step is simply a fixed linear functional of the sampled network
outputs, and therefore inherits the differentiability of $q(x;\theta)$ in
$\theta$: automatic differentiation propagates gradients through the
quadrature exactly as through any other layer of the computation, so
$\partial_\theta M_n^{\mathrm{pred}}$ is available directly from the same
graph used for $f(x;\theta)$, and no special-purpose implicit-differentiation
machinery is needed. In practice we use the composite Boole's rule, the
five-point closed Newton--Cotes quadrature~\cite{NR}, on a uniform mesh fine
enough that its truncation error is negligible against the resolution already
lost by working with a finite moment set; we do not reproduce the quadrature
formula here, as it plays a purely technical role.

Given target moments $\{M_n^{\mathrm{tar}}\}$, we match the moment estimates
just described with a weighted least-squares term,
\begin{equation}
  \mathcal{L}_{\mathrm{mom}}(\theta)=\sum_{n}\lambda_n\big(M_n^{\mathrm{pred}}(\theta)-M_n^{\mathrm{tar}}\big)^2,
  \label{eq:momloss}
\end{equation}
where the weights $\lambda_n$ are unity except for the first one or two orders,
which are enhanced (to $\sim\!10$) because they fix the global location and
width of the profile. The $n=0$ moment is already enforced by
Eq.~\eqref{eq:hardnorm} and is used only as a consistency check.

Because the inversion is ill posed, moment matching alone does not guarantee
the single-maximum, no-central-inflection behavior motivated in Sec.~2: in
exploratory runs it produced reconstructions with a second, spurious maximum,
or with a reversal of curvature near the dominant peak. We enforce the
preference argued for in Sec.~2 with two differentiable penalties on the
sampled values $q_i=q(x_i)$, evaluated on the same mesh used for quadrature.
The first enforces an ``increase-then-decrease'' (unimodal) profile by
penalizing the wrong sign of the discrete slope $\Delta q_i=q_{i+1}-q_i$ on
each side of the global-maximum index $k$,
\begin{equation}
  \mathcal{L}_{\mathrm{uni}}=
  \frac{1}{\max(1,k)}\sum_{i<k}\ReLU(-\Delta q_i)^2
  +\frac{1}{\max(1,M-k)}\sum_{i\ge k}\ReLU(\Delta q_i)^2 .
  \label{eq:uni}
\end{equation}
The second forbids inflection points in a central window around the peak (while
leaving the tails free) by penalizing positive curvature there,
\begin{equation}
  \mathcal{L}_{\mathrm{conc}}=\sum_{i:\,q_i\ge\tau q_{\max}}\ReLU\!\big(q''(x_i)\big)^2,
  \label{eq:conc}
\end{equation}
with $\tau\in(0,1)$ defining the central region by height. The full objective is
\begin{equation}
  \mathcal{L}(\theta)=\mathcal{L}_{\mathrm{mom}}
  +\lambda_{2}\,\mathcal{L}_{2}
  +\lambda_{\mathrm{uni}}\,\mathcal{L}_{\mathrm{uni}}
  +\lambda_{\mathrm{conc}}\,\mathcal{L}_{\mathrm{conc}},
  \label{eq:totalloss}
\end{equation}
where $\mathcal{L}_{2}=\sum_\ell\lVert A^{(\ell)}\rVert_F^2+\sum_\ell\lVert b^{(\ell)}\rVert_2^2$
is an $L_2$ penalty on every trainable parameter of the core, which stabilizes
training; for the KAN, the analogous sum runs over the base weights, spline
coefficients, and spline scalers of every edge, in place of $A^{(\ell)}$ and
$b^{(\ell)}$.

For the KAN alone we add one further term, absent for the MLP, which has no
counterpart in its parameterization. Writing $c^{(\ell)}_{ji,k}$ for the
$k$-th B-spline coefficient ($k=1,\dots,K$) of the edge from input $i$ to
output $j$ in layer $\ell$ (Sec.~3.2), define its mean absolute size
$\bar c^{(\ell)}_{ji}=\tfrac1K\sum_k|c^{(\ell)}_{ji,k}|$ and its normalized
weight within the layer, $p^{(\ell)}_{ji}=\bar c^{(\ell)}_{ji}\big/
\sum_{i',j'}\bar c^{(\ell)}_{j'i'}$. The KAN loss includes an additional
activation-sparsification penalty~\cite{KAN},
\begin{equation}
  \mathcal{L}_{\mathrm{spline}}
  =\sum_\ell\left[\,\sum_{i,j}\bar c^{(\ell)}_{ji}
  \;-\;\sum_{i,j}p^{(\ell)}_{ji}\log p^{(\ell)}_{ji}\,\right],
  \label{eq:splinereg}
\end{equation}
added to the total loss of Eq.~\eqref{eq:totalloss} with the same coefficient
$\lambda_2$ as $\mathcal{L}_2$. The first (entropy-free) term is an $L_1$
penalty that discourages large spline coefficients unless they materially
improve the fit; the second is the Shannon entropy of the normalized edge
magnitudes, which discourages spreading activation evenly across many edges.
Together they push the trained KAN toward a sparse, concentrated set of
active edges rather than a diffuse one---a standard KAN regularizer that has
no analogue for the MLP, whose parameters carry no such edge-wise structure
to sparsify.

A useful heuristic for the hyperparameters
$\lambda_{2},\lambda_{\mathrm{uni}},\lambda_{\mathrm{conc}}$ is to choose them so
that the loss contributions are of comparable magnitude in the first training
epochs.

\section{Results}
We benchmark the method on synthetic targets for which the exact moments and the
pointwise function are known~\cite{Xu:2024nzp}, so that both moment reproduction and pointwise
fidelity can be assessed. To quantify the variability induced by random
initialization, each reconstruction is reported as the mean over
$n_{\mathrm{ens}}$ independent trainings, with a band given by the pointwise
standard deviation.

\subsection{Symmetric benchmark (pion-like)}
\begin{figure}[t]
  \centering
  \includegraphics[width=0.48\textwidth]{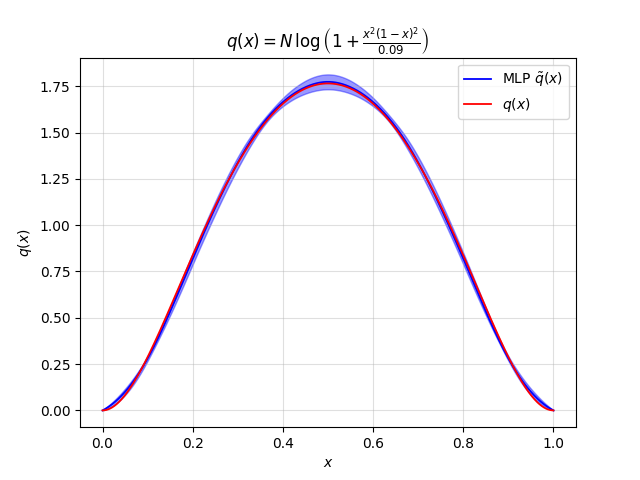}\hfill
  \includegraphics[width=0.48\textwidth]{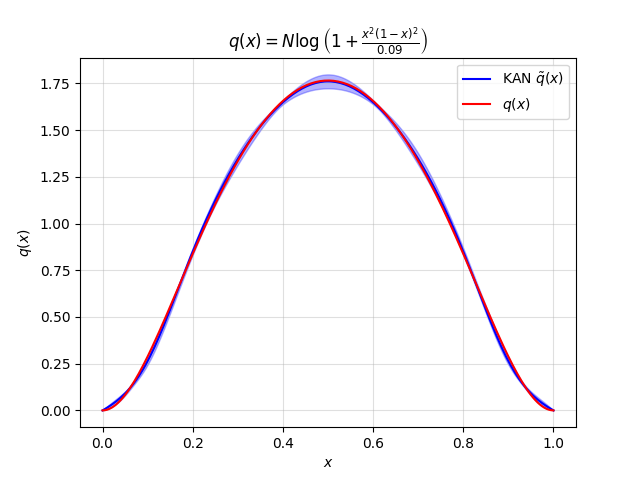}
  \caption{Symmetric benchmark, Eq.~\eqref{eq:qtest-sym} with $\rho=0.3$,
  trained on the first ten moments: MLP (left) and KAN (right)
  reconstructions, each shown as the ensemble mean ($n_{\mathrm{ens}}=30$)
  with a one-standard-deviation band.}
  \label{fig:sym}
\end{figure}

As a symmetric target we use the smooth function
\begin{equation}
  q_{\mathrm{test}}(x)=\mathcal{N}\,\log\!\Big(1+\frac{x^2(1-x)^2}{\rho^2}\Big),
  \qquad \rho=0.3,
  \label{eq:qtest-sym}
\end{equation}
with $\mathcal{N}$ fixed by unit normalization. The architecture enforces
$q(x)=q(1-x)$, endpoint vanishing, and normalization, and is trained on the
first $N=10$ moments. We compare the two cores of Sec.~3 under identical
constraints and loss: the MLP has a single hidden layer of $150$ neurons,
while the KAN has two hidden layers of $16$ nodes each, with spline order $3$
and grid size $10$; both are evaluated on a mesh of $301$ points. Each core is
trained independently $n_{\mathrm{ens}}=30$ times, and the reconstruction is
reported as the ensemble mean, with a band (figures) or parenthetical uncertainty
(tables) given by the pointwise standard deviation.
Table~\ref{tab:moments-sym} compares the exact and reconstructed moments for
both cores, and Fig.~\ref{fig:sym} shows the two pointwise reconstructions
side by side. We note a practical training detail: the Adam learning rate had
to be initialized at $10^{-5}$ for the first few hundred epochs and then
increased toward $10^{-3}$ to obtain stable convergence.

\begin{table}[t]
\centering
\caption{First ten Mellin moments for the symmetric benchmark,
Eq.~\eqref{eq:qtest-sym}. $M_n^{\mathrm{true}}$ are the exact moments;
$\overline{M}_n^{\mathrm{pred}}$ and
$\varepsilon_n=|\overline{M}_n^{\mathrm{pred}}-M_n^{\mathrm{true}}|/|M_n^{\mathrm{true}}|$
are the ensemble-averaged prediction and relative error, for the MLP and the
KAN (both $n_{\mathrm{ens}}=30$).}
\label{tab:moments-sym}
\vspace{0.25em}
\begin{tabular}{cccccc}
\toprule
& & \multicolumn{2}{c}{MLP} & \multicolumn{2}{c}{KAN} \\
\cmidrule(lr){3-4}\cmidrule(lr){5-6}
$n$ & $M_n^{\mathrm{true}}$ & $\overline{M}_n^{\mathrm{pred}}$ & $\varepsilon_n$
    & $\overline{M}_n^{\mathrm{pred}}$ & $\varepsilon_n$ \\
\midrule
1  & 0.5    & 0.4999997  & $7\times10^{-7}$ & 0.50000000(1) & $5\times10^{-9}$ \\
2  & 0.2884 & 0.2883(1)  & $3\times10^{-4}$ & 0.28838(6)    & $5\times10^{-5}$ \\
3  & 0.1826 & 0.1825(2)  & $7\times10^{-4}$ & 0.18257(8)    & $1\times10^{-4}$ \\
4  & 0.1234 & 0.1233(1)  & $7\times10^{-4}$ & 0.12341(6)    & $2\times10^{-4}$ \\
5  & 0.0876 & 0.08756(2) & $2\times10^{-4}$ & 0.08756(3)    & $2\times10^{-4}$ \\
6  & 0.0645 & 0.06457(9) & $9\times10^{-4}$ & 0.06450(5)    & $1\times10^{-4}$ \\
7  & 0.0489 & 0.0491(2)  & $3\times10^{-3}$ & 0.04895(9)    & $4\times10^{-5}$ \\
8  & 0.0381 & 0.0383(3)  & $5\times10^{-3}$ & 0.0381(1)     & $5\times10^{-4}$ \\
9  & 0.0302 & 0.0305(3)  & $9\times10^{-3}$ & 0.0302(2)     & $1\times10^{-3}$ \\
10 & 0.0244 & 0.0247(4)  & $1\times10^{-2}$ & 0.0244(2)     & $2\times10^{-3}$ \\
\bottomrule
\end{tabular}
\end{table}

\subsection{Asymmetric benchmark (kaon-like): MLP vs.\ KAN}
For the asymmetric target we use a Gegenbauer expansion built on the
$x^2(1-x)^2$ weight, which is the natural orthogonality weight of the
$C_j^{5/2}$ polynomials and encodes the Brodsky--Farrar endpoint behavior,
\begin{equation}
  q_{\mathrm{test}}(x)=\mathcal{N}\,x^2(1-x)^2\Big(1+\sum_{j=1}^{J}a_j\,C_j^{5/2}(1-2x)\Big),
  \label{eq:qtest-asym}
\end{equation}
where odd $a_j$ generate the asymmetry expected from SU(3) breaking. (This is a
valence-PDF-motivated construction, chosen for its endpoint behavior and
convenient orthogonal basis; it should not be confused with the $C_n^{3/2}$
ERBL eigenbasis of Sec.~2, which applies to the PDA.) We consider two
instances of Eq.~\eqref{eq:qtest-asym}, both fit from the first $N=10$
moments: a two-term expansion with $(a_1,a_2)=(0.1,0.02)$, and a four-term
expansion with $(a_1,a_2,a_3,a_4)=(-0.08,0.1,-0.01,0.013)$, the latter of
similar magnitude to (and rounded from) the Gegenbauer-moment values reported
for the kaon distribution amplitude in Table~I of Ref.~\cite{Raya}. We
emphasize that these coefficients are used here only as a representative,
realistically sized asymmetric input for the $C_j^{5/2}$, valence-PDF-style
test function of Eq.~\eqref{eq:qtest-asym}, not as a literal instance of the
kaon PDA itself, whose natural expansion is the distinct $C_n^{3/2}$ basis
discussed above. Both instances are fit with the same architectures and
training setup: the MLP has three hidden layers of $8$ neurons each, and the
KAN has the same architecture as in Sec.~5.1 (two hidden layers of $16$ nodes,
spline order $3$, grid size $10$); each core is trained independently
$n_{\mathrm{ens}}=30$ times per instance.

\subsubsection*{Two-term expansion, $(a_1,a_2)=(0.1,0.02)$}
Table~\ref{tab:asym1} compares the exact and reconstructed moments for both
cores, and the top row of Fig.~\ref{fig:asym} shows the pointwise
reconstructions.

\begin{table}[t]
\centering
\caption{Asymmetric benchmark, Eq.~\eqref{eq:qtest-asym} with
$(a_1,a_2)=(0.1,0.02)$. $M_n^{\mathrm{true}}$, $\overline{M}_n^{\mathrm{pred}}$,
and $\varepsilon_n=|\overline{M}_n^{\mathrm{pred}}-M_n^{\mathrm{true}}|/|M_n^{\mathrm{true}}|$
are defined as in Table~\ref{tab:moments-sym}, for the MLP and the KAN
($n_{\mathrm{ens}}=30$ each).}
\label{tab:asym1}
\vspace{0.25em}
\begin{tabular}{cccccc}
\toprule
& & \multicolumn{2}{c}{MLP} & \multicolumn{2}{c}{KAN} \\
\cmidrule(lr){3-4}\cmidrule(lr){5-6}
$n$ & $M_n^{\mathrm{true}}$ & $\overline{M}_n^{\mathrm{pred}}$ & $\varepsilon_n$
    & $\overline{M}_n^{\mathrm{pred}}$ & $\varepsilon_n$ \\
\midrule
1  & 0.4643 & 0.4640(1)  & $5\times10^{-4}$ & 0.4643(3) & $5\times10^{-5}$ \\
2  & 0.2524 & 0.2527(1)  & $1\times10^{-3}$ & 0.2525(4) & $4\times10^{-4}$ \\
3  & 0.1524 & 0.1525(1)  & $8\times10^{-4}$ & 0.1523(3) & $6\times10^{-4}$ \\
4  & 0.0991 & 0.0991(1)  & $8\times10^{-4}$ & 0.0989(3) & $2\times10^{-3}$ \\
5  & 0.0682 & 0.06803(7) & $2\times10^{-3}$ & 0.0680(3) & $3\times10^{-3}$ \\
6  & 0.0490 & 0.04881(5) & $3\times10^{-3}$ & 0.0488(2) & $3\times10^{-3}$ \\
7  & 0.0364 & 0.03628(5) & $2\times10^{-3}$ & 0.0363(2) & $1\times10^{-3}$ \\
8  & 0.0278 & 0.02775(7) & $7\times10^{-4}$ & 0.0278(2) & $2\times10^{-3}$ \\
9  & 0.0217 & 0.0218(1)  & $2\times10^{-3}$ & 0.0218(1) & $6\times10^{-3}$ \\
10 & 0.0173 & 0.0174(1)  & $6\times10^{-3}$ & 0.0175(1) & $1\times10^{-2}$ \\
\bottomrule
\end{tabular}
\end{table}

\subsubsection*{Four-term expansion, $(a_1,a_2,a_3,a_4)=(-0.08,0.1,-0.01,0.013)$}
\begin{figure}[t]
  \centering
  \includegraphics[width=0.48\textwidth]{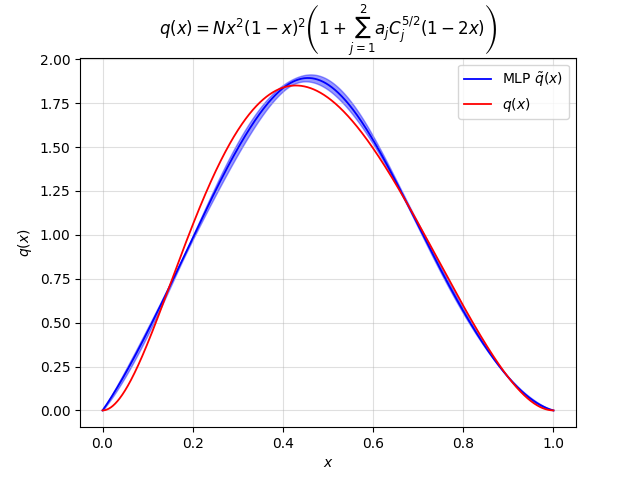}\hfill
  \includegraphics[width=0.48\textwidth]{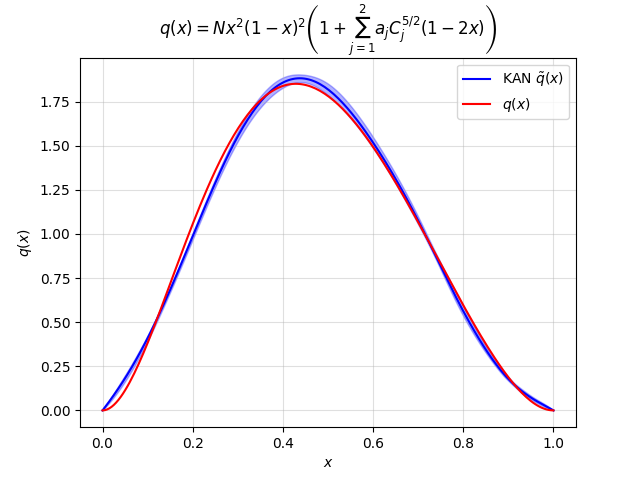}\\[0.5em]
  \includegraphics[width=0.48\textwidth]{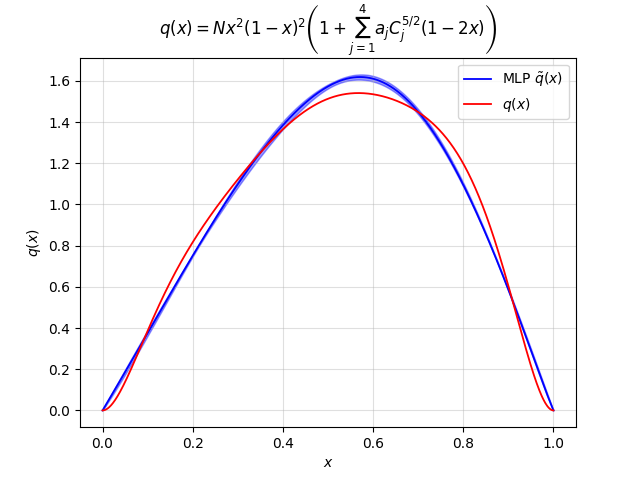}\hfill
  \includegraphics[width=0.48\textwidth]{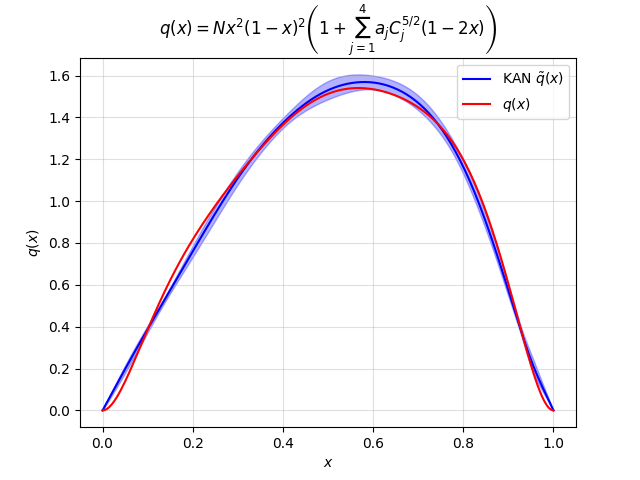}
  \caption{Asymmetric (kaon-like) benchmark, Eq.~\eqref{eq:qtest-asym}: MLP
  (left column) and KAN (right column) reconstructions against the exact
  distribution, for the two-term expansion $(a_1,a_2)=(0.1,0.02)$ (top row)
  and the four-term expansion
  $(a_1,a_2,a_3,a_4)=(-0.08,0.1,-0.01,0.013)$ adapted from Ref.~\cite{Raya} (bottom
  row). Each panel shows the ensemble mean ($n_{\mathrm{ens}}=30$) with a
  one-standard-deviation band.}
  \label{fig:asym}
\end{figure}

Table~\ref{tab:asym2} compares the exact and reconstructed moments for both
cores, and the bottom row of Fig.~\ref{fig:asym} shows the pointwise
reconstructions. This instance is more structured than the two-term
one, and moment reproduction degrades faster with $n$ for both cores, but
markedly less so for the KAN.

\begin{table}[t]
\centering
\caption{Asymmetric benchmark, Eq.~\eqref{eq:qtest-asym} with
$(a_1,a_2,a_3,a_4)=(-0.08,0.1,-0.01,0.013)$, adapted from Ref.~\cite{Raya}.
$M_n^{\mathrm{true}}$, $\overline{M}_n^{\mathrm{pred}}$, and $\varepsilon_n$
as in Table~\ref{tab:asym1}.}
\label{tab:asym2}
\vspace{0.25em}
\begin{tabular}{cccccc}
\toprule
& & \multicolumn{2}{c}{MLP} & \multicolumn{2}{c}{KAN} \\
\cmidrule(lr){3-4}\cmidrule(lr){5-6}
$n$ & $M_n^{\mathrm{true}}$ & $\overline{M}_n^{\mathrm{pred}}$ & $\varepsilon_n$
    & $\overline{M}_n^{\mathrm{pred}}$ & $\varepsilon_n$ \\
\midrule
1  & 0.5286 & 0.5288(1)  & $4\times10^{-4}$ & 0.5286(2) & $2\times10^{-5}$ \\
2  & 0.3262 & 0.3259(1)  & $9\times10^{-4}$ & 0.3262(2) & $3\times10^{-5}$ \\
3  & 0.2206 & 0.2200(1)  & $3\times10^{-3}$ & 0.2204(2) & $9\times10^{-4}$ \\
4  & 0.1585 & 0.15799(5) & $3\times10^{-3}$ & 0.1582(2) & $2\times10^{-3}$ \\
5  & 0.1188 & 0.11861(8) & $2\times10^{-3}$ & 0.1186(1) & $2\times10^{-3}$ \\
6  & 0.0920 & 0.0921(2)  & $2\times10^{-3}$ & 0.0919(1) & $1\times10^{-3}$ \\
7  & 0.0730 & 0.0735(2)  & $7\times10^{-3}$ & 0.0730(1) & $4\times10^{-4}$ \\
8  & 0.0591 & 0.0599(2)  & $1\times10^{-2}$ & 0.0593(2) & $3\times10^{-3}$ \\
9  & 0.0486 & 0.0497(3)  & $2\times10^{-2}$ & 0.0489(3) & $6\times10^{-3}$ \\
10 & 0.0406 & 0.0419(3)  & $3\times10^{-2}$ & 0.0410(3) & $1\times10^{-2}$ \\
\bottomrule
\end{tabular}
\end{table}

Across both instances the KAN reconstructs the pointwise profile of the
target distribution more faithfully than the MLP, consistent with the
symmetric benchmark of Sec.~5.1. We stress that this advantage refers to the
$x$-profile, not to each individual moment: both cores reproduce the input
moments at a comparable level (in Table~\ref{tab:asym1} the MLP is even
slightly more precise for several $n$), which is precisely the ill-posedness
discussed in Sec.~2---nearly identical moment sets are compatible with
visibly different profiles, as the MLP panels of Fig.~\ref{fig:asym} show.
What does distinguish the KAN systematically is the lowest orders: in all
three benchmarks it reproduces the moments $n=1$--$3$ more precisely than the
MLP, by up to an order of magnitude (Tables~\ref{tab:moments-sym},
\ref{tab:asym1}, and~\ref{tab:asym2}). These are the moments that fix the
location, width, and skewness of the profile, and which the loss already
weights most heavily, so their more accurate reproduction is consistent with
the KAN's more faithful pointwise reconstruction---although, the inversion
being ill posed, low-order moment accuracy alone cannot guarantee it.
At the level of moments, the difference is most visible in the higher moments
of the four-term, more structured instance, where the MLP's relative
error grows to $3\times10^{-2}$ at $n=10$ against $1\times10^{-2}$ for the KAN.

\subsection{Effect of the number of input moments (KAN)}
Having established that the KAN outperforms the MLP at fixed moment count, we
now assess, for the KAN alone, how the reconstruction degrades as fewer
moments are supplied. (The MLP was also trained on $N=5$ moments; its
reconstructions deviated from the true profile considerably more than the
KAN ones, and are not shown.) We repeat both asymmetric instances of the previous
section, training the same KAN architecture on only the first $N=5$ moments,
again with $n_{\mathrm{ens}}=30$ independent trainings, and compare the
resulting reconstruction to the $N=10$ case already reported, both pointwise
and for all ten moments---including the five beyond what the $N=5$ fit was
ever asked to match.

\begin{figure}[t]
  \centering
  \includegraphics[width=0.48\textwidth]{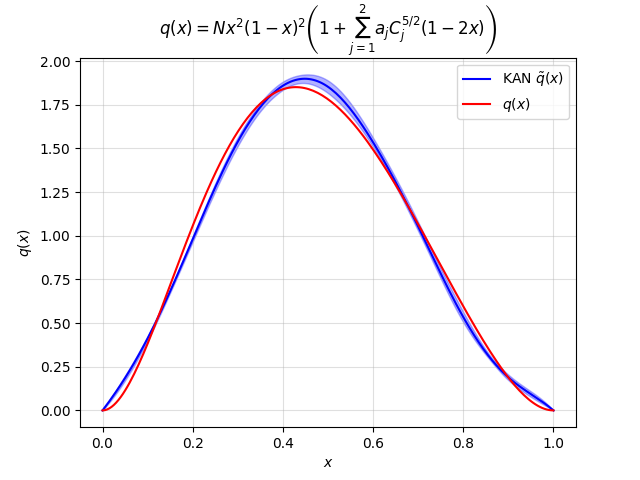}\hfill
  \includegraphics[width=0.48\textwidth]{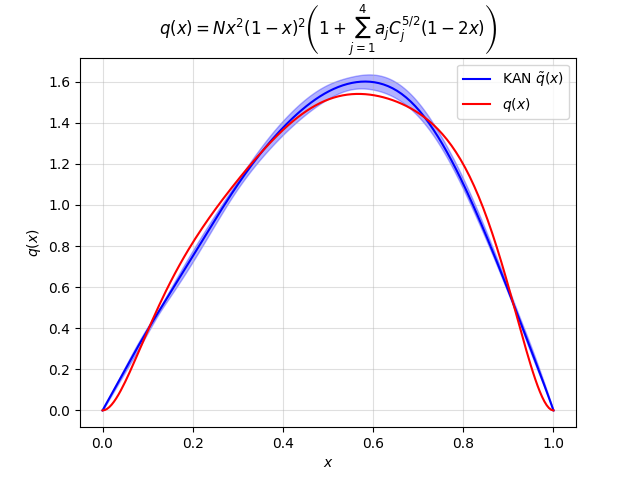}
  \caption{KAN reconstructions using only the first $N=5$ moments, for the
  two-term expansion $(a_1,a_2)=(0.1,0.02)$ (left) and the four-term expansion
  $(a_1,a_2,a_3,a_4)=(-0.08,0.1,-0.01,0.013)$ adapted from Ref.~\cite{Raya} (right).
  Each panel shows the ensemble mean ($n_{\mathrm{ens}}=30$) with a
  one-standard-deviation band; compare to the $N=10$ KAN panels (right
  column) of Fig.~\ref{fig:asym}.}
  \label{fig:asym-5mom}
\end{figure}

Tables~\ref{tab:kan-nmom-1} and~\ref{tab:kan-nmom-2} compare the $N=10$ and
$N=5$ KAN moments for the two-term and four-term expansions, respectively, and
Fig.~\ref{fig:asym-5mom} shows the corresponding $N=5$ pointwise
reconstructions. In both instances the moments that entered the $N=5$
training ($n\le5$, above the rule in each table) remain about as well
reproduced as in the $N=10$ fit, but the higher moments ($n>5$, below the
rule), which the $N=5$ network never saw during training, show visibly larger
deviations from the true values than their $N=10$ counterparts---most
markedly at $n=9,10$, where the relative error grows by roughly a factor of
four to five. This is the expected signature of a more strongly ill-posed
inversion at smaller $N$: with less information about the tail behavior
supplied during training, the reconstruction is less constrained away from
the region directly probed by the low moments, and the pointwise bands in
Fig.~\ref{fig:asym-5mom} are correspondingly wider than the $N=10$ KAN panels
(right column) of Fig.~\ref{fig:asym}.

\begin{table}[t]
\centering
\caption{KAN reconstruction of the two-term asymmetric benchmark,
Eq.~\eqref{eq:qtest-asym} with $(a_1,a_2)=(0.1,0.02)$, trained on the first
$N=10$ vs.\ the first $N=5$ moments ($n_{\mathrm{ens}}=30$ each).
$M_n^{\mathrm{true}}$, $\overline{M}_n^{\mathrm{pred}}$, and $\varepsilon_n$ as
in Table~\ref{tab:moments-sym}. The rule after $n=5$ separates the moments
used to train the $N=5$ model from those extrapolated beyond it.}
\label{tab:kan-nmom-1}
\vspace{0.25em}
\begin{tabular}{cccccc}
\toprule
& & \multicolumn{2}{c}{KAN, $N=10$} & \multicolumn{2}{c}{KAN, $N=5$} \\
\cmidrule(lr){3-4}\cmidrule(lr){5-6}
$n$ & $M_n^{\mathrm{true}}$ & $\overline{M}_n^{\mathrm{pred}}$ & $\varepsilon_n$
    & $\overline{M}_n^{\mathrm{pred}}$ & $\varepsilon_n$ \\
\midrule
1  & 0.4643 & 0.4643(3) & $5\times10^{-5}$ & 0.4642(2) & $2\times10^{-4}$ \\
2  & 0.2524 & 0.2525(4) & $4\times10^{-4}$ & 0.2523(3) & $2\times10^{-4}$ \\
3  & 0.1524 & 0.1523(3) & $6\times10^{-4}$ & 0.1522(2) & $1\times10^{-3}$ \\
4  & 0.0991 & 0.0989(3) & $2\times10^{-3}$ & 0.0989(2) & $2\times10^{-3}$ \\
5  & 0.0682 & 0.0680(3) & $3\times10^{-3}$ & 0.0681(2) & $7\times10^{-4}$ \\
\midrule
6  & 0.0490 & 0.0488(2) & $3\times10^{-3}$ & 0.0491(2) & $3\times10^{-3}$ \\
7  & 0.0364 & 0.0363(2) & $1\times10^{-3}$ & 0.0367(3) & $9\times10^{-3}$ \\
8  & 0.0278 & 0.0278(2) & $2\times10^{-3}$ & 0.0283(3) & $2\times10^{-2}$ \\
9  & 0.0217 & 0.0218(1) & $6\times10^{-3}$ & 0.0224(4) & $3\times10^{-2}$ \\
10 & 0.0173 & 0.0175(1) & $1\times10^{-2}$ & 0.0181(4) & $5\times10^{-2}$ \\
\bottomrule
\end{tabular}
\end{table}

\begin{table}[t]
\centering
\caption{KAN reconstruction of the four-term asymmetric benchmark,
Eq.~\eqref{eq:qtest-asym} with $(a_1,a_2,a_3,a_4)=(-0.08,0.1,-0.01,0.013)$,
adapted from Ref.~\cite{Raya}, trained on the first $N=10$ vs.\ the first $N=5$
moments ($n_{\mathrm{ens}}=30$ each). $M_n^{\mathrm{true}}$,
$\overline{M}_n^{\mathrm{pred}}$, and $\varepsilon_n$ as in
Table~\ref{tab:moments-sym}; the rule after $n=5$ is as in
Table~\ref{tab:kan-nmom-1}.}
\label{tab:kan-nmom-2}
\vspace{0.25em}
\begin{tabular}{cccccc}
\toprule
& & \multicolumn{2}{c}{KAN, $N=10$} & \multicolumn{2}{c}{KAN, $N=5$} \\
\cmidrule(lr){3-4}\cmidrule(lr){5-6}
$n$ & $M_n^{\mathrm{true}}$ & $\overline{M}_n^{\mathrm{pred}}$ & $\varepsilon_n$
    & $\overline{M}_n^{\mathrm{pred}}$ & $\varepsilon_n$ \\
\midrule
1  & 0.5286 & 0.5286(2)  & $2\times10^{-5}$ & 0.52860(6) & $5\times10^{-5}$ \\
2  & 0.3262 & 0.3262(2)  & $3\times10^{-5}$ & 0.32616(8) & $9\times10^{-5}$ \\
3  & 0.2206 & 0.2204(2)  & $9\times10^{-4}$ & 0.22050(5) & $5\times10^{-4}$ \\
4  & 0.1585 & 0.1582(2)  & $2\times10^{-3}$ & 0.1585(1)  & $3\times10^{-5}$ \\
5  & 0.1188 & 0.1186(1)  & $2\times10^{-3}$ & 0.1191(2)  & $2\times10^{-3}$ \\
\midrule
6  & 0.0920 & 0.0919(1)  & $1\times10^{-3}$ & 0.0926(3)  & $6\times10^{-3}$ \\
7  & 0.0730 & 0.0730(1)  & $4\times10^{-4}$ & 0.0739(4)  & $1\times10^{-2}$ \\
8  & 0.0591 & 0.0593(2)  & $3\times10^{-3}$ & 0.0603(4)  & $2\times10^{-2}$ \\
9  & 0.0486 & 0.0489(3)  & $6\times10^{-3}$ & 0.0501(4)  & $3\times10^{-2}$ \\
10 & 0.0406 & 0.0410(3)  & $1\times10^{-2}$ & 0.0422(4)  & $4\times10^{-2}$ \\
\bottomrule
\end{tabular}
\end{table}

\FloatBarrier

\section{Discussion and outlook}
Our aim has been to reconstruct meson distributions from finite moment data with
minimal functional bias: a flexible neural core is shown to accurately capture the shape, while only
first-principles properties---support, endpoint behavior, normalization, and (for
the pion and quarkonia) the $x\leftrightarrow 1-x$ symmetry---are hard-wired, and soft penalties
suppress the multiple-maxima and central-concavity artifacts of ill-posed
inversion. Replacing the MLP core by a KAN improves the accuracy with which the
pointwise profile is reconstructed, most visibly for the asymmetric kaon and at the smallest moment counts;
we attribute this to the compact support of the B-spline activations, which
localizes moment-driven corrections and damps the global oscillations to which
the MLP is susceptible.

This comparison is not matched in parameter count: for the architectures of
Sec.~5, the KAN carries substantially more trainable parameters than the MLP
(of order $4\times10^3$ for the KAN cores used here, against a few hundred for
the MLP, i.e.\ roughly an order of magnitude more in the symmetric benchmark
and over twenty-fold more in the asymmetric one), so capacity alone could in
principle account for part of the KAN's advantage. We do not believe this is
the explanation, however: enlarging the MLP core, by widening or adding hidden
layers, did not improve its reconstructions in our trials and in fact made
them less stable, consistent with the picture of Sec.~3.2---additional
capacity gives an architecture with globally supported activations more ways
to fit the target moments with a spurious oscillation, not fewer. A
controlled comparison at matched parameter count, isolating the effect of the
spline basis from raw capacity, would nonetheless make this conclusion more
rigorous and is a natural next step.

Furthermore, having demonstrated that the neural cores described here can
reproduce the shape of a distribution from a limited set of moments, imposing
only a few well-grounded physical constraints and no particular functional
form, we plan to follow this feasibility study with a reconstruction of meson
PDFs and PDAs from the moments available in the literature, as obtained within
the DSE/BSE framework.

On the other hand, a limitation of this ``fit-per-dataset'' strategy is that a full optimization is
required for every new moment sequence, so it is not an instantaneous emulator.
A natural complement, borrowed from the variational- and
simulation-based-inference literature~\cite{KingmaWelling,CranmerBrehmerLouppe},
is \emph{amortized inference}: rather than solving a fresh optimization problem
for every new instance, as we do here, a single network is trained once on a
large collection of instances, so that the cost of inference is paid up front
during training and ``amortized'' over all subsequent uses---at test time a new
instance requires only a forward pass through the trained network, not a new
optimization. In our setting this means training a single model on a large
synthetic library of moment sequences drawn from many functional families,
mapping the (sequential) input $\{M_n\}_{n=1}^{N}$ directly to the reconstructed
mesh values $\{q(x_i)\}$. The sequential structure is naturally suited to a
Transformer. Note that this supervised setting does not admit the hard/soft
constraints of Secs.~2--4 in the same form: the network regresses directly on
mesh values rather than on an unconstrained core that is subsequently wrapped
or penalized, so there is neither an architectural wrapper to apply nor a
moment-matching loss to augment with shape penalties. The constraints are
instead retained only implicitly, through the choice of training data: the
functional families used to generate the synthetic library are themselves
built to satisfy support, endpoint, symmetry, and unimodality properties by
construction, so the network only ever learns from (and hence only ever
outputs) constraint-satisfying profiles. We have already obtained encouraging
results along these lines with an
encoder-only Transformer architecture, which we plan to present in a
forthcoming paper. Such an approach trades some functional
bias---bounded by the diversity of the training library---for immediate
inference once trained, enabling fast scans and uncertainty propagation. The
two strategies are thus complementary: the present constrained fit minimizes
bias per dataset, while the amortized model prioritizes rapid inference after
an up-front training cost.

\appendix
\section{B-splines}
\label{app:bsplines}
This appendix collects the definitions needed to make the KAN construction of
Sec.~3.2 self-contained; see~\cite{deBoor} for a complete treatment. Fix a
non-decreasing sequence of \emph{knots} $t_0\le t_1\le\cdots\le t_{m}$ on the
domain of interest. The B-splines of degree $0$ associated with this knot
sequence are the piecewise-constant indicator functions of the knot intervals,
\begin{equation}
  B_{i,0}(x) =
  \begin{cases}
    1, & t_i \le x < t_{i+1}, \\
    0, & \text{otherwise},
  \end{cases}
  \label{eq:bspline-deg0}
\end{equation}
and the B-splines of degree $p\ge1$ are defined from those of degree $p-1$ by
the Cox--de Boor recursion,
\begin{equation}
  B_{i,p}(x) =
  \frac{x-t_i}{t_{i+p}-t_i}\,B_{i,p-1}(x)
  \;+\;
  \frac{t_{i+p+1}-x}{t_{i+p+1}-t_{i+1}}\,B_{i+1,p-1}(x),
  \label{eq:cox-deboor}
\end{equation}
with the convention that any term with a vanishing denominator is set to zero.
Each $B_{i,p}$ is itself a spline of degree $p$: a polynomial of degree $p$ on
every sub-interval $[t_j,t_{j+1})$, joined $p-1$ times continuously
differentiably across knots. Iterating Eq.~\eqref{eq:cox-deboor} shows that
$B_{i,p}$ is supported on $[t_i,t_{i+p+1}]$ and vanishes identically outside
this window of $p+1$ consecutive knot intervals---the compact-support property
used in Sec.~3.2. Any spline of degree $p$ on the given knot sequence can be
written uniquely as a linear combination
\begin{equation}
  s(x)=\sum_i c_i\,B_{i,p}(x),
\end{equation}
and, because of the support property above, changing a single coefficient
$c_i$ leaves $s(x)$ unchanged everywhere outside $[t_i,t_{i+p+1}]$. This is the
property Eq.~\eqref{eq:kan-edge} relies on, with $p$ fixed by the spline order
of the KAN and $\{t_i\}$ a fixed, uniform knot grid on the edge's input range.

\section*{Acknowledgments}
Work supported by the Spanish Ministry of Science and Innovation (MICINN grant no.\ PID2022-140440NB-C22, PID2025-171916NB-C32), and Junta de Andaluc\'ia (grant no.\ P18-FR-5057). KR acknowledges the Physics Department at Cinvestav-Zacatenco for its hospitality and support.

\bibliographystyle{unsrt}
\bibliography{bibliography}

\end{document}